\documentclass[traditabstract]{aa}
\usepackage{graphicx}
\usepackage{rotating,subfigure,amssymb,afterpage}
\usepackage{txfonts}
\usepackage{natbib}
\begin{document}
   \title{The Cosmic Large-Scale Structure in X-rays (CLASSIX) \\
   Cluster Survey II:
Unveiling a pancake structure with a 100 Mpc radius in the local Universe
\thanks{
   Based on observations at the European Southern Observatory La Silla,
   Chile and the German-Spanish Observatory at Calar Alto}}

   \author{Hans B\"ohringer\inst{1,2}, Gayoung Chon\inst{1}, Joachim Tr\"umper\inst{2}} 

   \offprints{H. B\"ohringer, hxb@mpe.mpg.de}
   \institute{$^1$ Universit\"ats-Sternwarte M\"unchen, Fakult\"at f\"ur Physik,
                  Ludwig-Maximilians-Universit\"at M\"unchen,
                  Scheinerstr. 1, 81679 M\"unchen, Germany.\\
              $^2$ Max-Planck-Institut f\"ur extraterrestrische Physik,
                   D-85748 Garching, Germany.
}

   \date{Submitted 17/02/21}

\abstract{Previous studies of the galaxy and galaxy cluster distribution in the local
Universe found indications for a large extension of the Local Supercluster up to
a radius of $190 h_{70}^{-1}$ Mpc. 
We are using our large and highly complete {\sf CLASSIX} survey of X-ray luminous 
galaxy clusters detected in the ROSAT All Sky Survey to trace the matter distribution in
the local Universe and to explore the size of the flattened local density structure associated with
the Local Supercluster. The Local Supercluster is oriented almost perpendicular to the
Galactic plane. Since Galactic extinction increases towards the Galactic plane,
objects are on average more easily visible perpendicular to the plane than close to it, 
also producing an apparent concentration of objects along the Local Supercluster.
We can correct for this bias by a careful treatment of the survey selection function. 
We find a significant overdensity of clusters in a flattened structure along the Supergalactic 
plane with a thickness of about 50 Mpc and an extent of about 100 Mpc radius. Structures
at a distance larger than 100 Mpc are not correlated to the Local Supercluster any more.
The matter density contrast of the local superstructure to the surroundings is about a factor 
of 1.3 - 2.3. Within the Supergalactic plane the matter is concentrated mostly in two
superclusters, the Perseus-Pisces Chain and Hydra-Centaurus supercluster.
We have shown in our earlier work that the local Universe in a region with a radius of 
100 - 170 Mpc has a lower density than the cosmic mean. For this reason, the Local Supercluster
is not overdense with respect to the cosmic mean density. Therefore this local superstructure
will not collapse as a whole in the future, but rather fragment.
}


\authorrunning{B\"ohringer et al.}
\titlerunning{The Local Pancake Superstructure}
   \maketitle
%

\section{Introduction}

Our local neighbourhood in the Universe is characterised by a flattened matter
density distribution which seems to show
coherence over a hierarchy of scales. Studying the density distribution of
bright galaxies in the sky, notably those compiled in the Shapley-Ames galaxy
survey, De Vaucouleurs noted a concentration of these galaxies towards
a plane roughly perpendicular to the plane of our Galaxy 
\citep{Dev1953,Dev1956,Dev1958,Dev1959}.
He called this pronounced structure the `Local Supergalactic System', 
which is now referred to as the Local Supercluster. The system
includes a number of galaxy groups and the Virgo cluster as its dominant
member. This concentration of galaxies was noted earlier
by William Herschel (see \citet{Fli1986} and \citet{Rub1951} for a historical 
account).

Already at much smaller scales, there is a pronounced segregation of
galaxies towards the Supergalactic plane. The analysis of the galaxy 
distribution within a radius of about 6.5 Mpc by \citet{Cou2013}, 
for example, shows that almost all
the nearby groups of galaxies, including the Local Group, the groups marked by 
IC342, M81, Cen A, M83, N4214, and the Canes Venatici I and Maffei I groups,
are contained in a very narrow slice, which is aligned with the Local Supercluster 
(see their Fig. 3). Other studies have found that this flattened
structure also reaches far beyond the Local Supercluster. For example,
\citet{Sha1999} found in the distribution of nearby
bright radio galaxies an indication of a concentration 
of these objects towards the Supergalactic Plane with an extent up to 
about a redshift of $ z \sim 0.02$ ($ \sim 85.3 h_{70}^{-1}$ Mpc). 
Investigating the distribution
of 48 nearby Abell clusters, \citet{Tul1986,Tul1987} described a concentration
which extends the Local Supercluster to a flattened superstructure 
with a diameter of about  360 $h_{75}^{-1}$ Mpc, stating that the orientation
of the short axis is coinciding with the pole of the plane of the traditional
Local Supercluster. 

One problem for a quantitative mapping of the local superstructure is, however,
the fact that observations are affected by Galactic absorption and extinction,
which depend strongly on Galactic latitude. Since the Supergalactic plane is
almost perpendicular to the Galactic disk, the Supergalactic poles are not 
affected by strong extinction while the Supergalactic plane is partly shaded 
by the Galactic band. This can lead to
an apparent enhancement of the galaxy concentration towards the Supergalactic
plane. We therefore reapproached the mapping of the local superstructure
with tracer objects, for which such sampling bias can be avoided or properly corrected
for.
 
For this reason, we used galaxy clusters to trace the large-scale structure.
As an integral part of the cosmic large-scale structure, galaxy clusters
are reliable tracers of the underlying dark matter distribution. Since they 
form from the largest peaks in the initially random Gaussian density fluctuation 
field, their density distribution can be statistically closely related 
to the matter density distribution (e.g. \cite{Bar1986}). Cosmic
structure formation theory has shown that the ratio of the cluster density
fluctuation amplitude is enhanced with respect to the matter density fluctuations
in the sense that the cluster density fluctuations follow the matter
density fluctuations with a higher amplitude. The ratio of the amplitude 
of the cluster density to that of the dark
matter is practically independent of the scale \citep{Kai1986,Mo1996,She1999}.
The fact that the cluster density fluctuation amplitude is enhanced makes
clusters sensitive tracers of the matter distribution.

We used the large, highly complete sample of X-ray luminous galaxy
clusters from the {\sf CLASSIX} (Cosmic Large-Scale Structure in X-rays) galaxy 
cluster survey to probe the matter distribution in the nearby Universe.
The detection of the objects
in X-rays provides the advantage of X-ray emission ensuring that the systems
are tight three-dimensional mass concentrations, and the X-ray luminosity provides
a good estimate of the object's mass. Therefore, by construction, an X-ray flux 
limited survey is practically mass selected (with some scatter)
with a well-defined mass limit as a function of redshift. These properties of the
survey are essential for the goal of this study. We have already applied 
{\sf CLASSIX} to successful studies of the cosmic large-scale structure as described
in section 2 and the present study can build on this experience.

The paper is structured as follows. In section 2 we describe the construction and
properties of the {\sf CLASSIX} galaxy cluster survey and some relevant applications.
Section 3 deals with some methodological issues. The results of our analysis is
presented in section 4 and implications are discussed in section 5. Section
6 provides a summary and conclusions.
For physical properties, which depend on distance, we use the following
cosmological parameters: a Hubble constant of $H_0 = 70$ km s$^{-1}$ Mpc$^{-1}$,
a matter density, $\Omega_m = 0.3$, and a spatially flat metric. 
For the cosmographical analysis, we
use Supergalactic coordinates, defined by the location of the
Supergalactic North Pole at $l_{II} = 47.3700$~deg and $b_{II} = 6.3200$ deg,
consistent with the definition by De Vaucouleurs et al. in the Third Catalog 
of Bright Galaxies (1991, see also \cite{Lah2000}).

\section{The CLASSIX galaxy cluster survey}

This study requires a cluster catalogue that traces the local Universe  
densely enough, is statistically highly complete, and has a 
well-known selection function.
At the moment, the best data base is the {\sf CLASSIX} galaxy cluster catalogue
\citep{Boe2016}.
It is the combination of our surveys in the southern sky,  {\sf REFLEX II}
\citep{Boe2013}, and the northern hemisphere, {\sf NORAS II}
\citep{Boe2017}. Together they cover 8.26 ster of the sky
at galactic latitudes $|b_{II}| \ge 20^o$
and the cluster catalogue contains 1773 members.
In this study, we do not excise the regions of the Magellanic Clouds
or the VIRGO cluster. In the complete survey, we find no
significant deficit in the cluster density in these sky areas.
We also used an extension of CLASSIX to lower galactic latitudes into the
`zone of avoidance' (ZoA). This region is restricted to the area with an
interstellar hydrogen column density $n_H \le 2.5 \times 10^{21}$ cm$^{-2}$,
because in regions with a higher column density, X-rays are strongly absorbed
and usually the sky has a high stellar density, making the detection
of clusters in the optical extremely difficult. The values for the
interstellar hydrogen column density are taken from the 21cm survey of
\citet{Dic1990}~\footnote{We have compared the
interstellar hydrogen column density compilation by \citet{Dic1990}
with the more recent data set of the 
Bonn-Leiden-Argentine 21cm survey \citep{Kal2005}
and found that the differences relevant for us are of the order of at most one percent.
Because our survey was constructed with a flux cut based
on the Dickey \& Lockman results, we kept the older hydrogen column density
values for consistency reasons.}.
This area amounts to another
2.56 ster and altogether the survey data cover 86.2\% of the sky. The spectroscopic
follow-up to obtain redshifts for this part of the survey is about 70\%
complete and also the completeness of the cluster sample is not as high as for 
{\sf REFLEX} and {\sf NORAS}. The cluster density we used for the 
ZoA is therefore a lower limit. In total, we have 143 galaxy clusters with a
redshift $z \le 0.03$ and an X-ray luminosity $L_X \ge 10^{42}$ erg s$^{-1}$
that we use in this study.

The {\sf CLASSIX} galaxy cluster survey and its extension is based on 
the X-ray detection of galaxy clusters in the ROSAT All-Sky Survey
(RASS, \cite{Tru1993,Vog1999}). The source detection for the survey, 
the construction of the survey, and 
the survey selection function  as well as tests of the completeness of the
survey are described in \citet{Boe2013, Boe2017}. In summary, the 
nominal unabsorbed flux limit for the galaxy cluster detection in the RASS is
$1.8 \times 10^{-12}$ erg s$^{-1}$ cm$^{-2}$ in the
0.1 - 2.4 keV energy band. For the assessment of the large-scale structure
in this paper, we apply an additional cut
on the minimum number of detected source photons of 20 counts. This has
the effect that the nominal flux limit quoted above is only reached in about
80\% of the survey. In regions with lower exposure and higher interstellar
absorption, the flux limit is accordingly higher 
(see Fig.\ 11 in \citet{Boe2013} and Fig.\ 5 in \citet{Boe2017}). 
This effect is modelled and
taken into account in the survey selection function.

We have already demonstrated with the {\sf REFLEX I} survey 
\citep{Boe2004} that clusters provide a precise means to 
obtain a census of the cosmic large-scale matter distribution
through, for example, the correlation function \citep{Col2000}, 
the power spectrum \citep{Sch2001, Sch2002, Sch2003a, Sch2003b}, 
Minkowski functionals \citep{Ker2001},
and, using {\sf REFLEX II}, with the study
of superclusters \citep{Cho2013,Cho2014} and the cluster power 
spectrum \citep{Bal2011,Bal2012}. 
The latter study also demonstrates that the theoretically predicted
large-scale structure bias, that is the amplification factor, with which
the clusters trace the matter density fluctuations is confirmed
by observations.

Relevant physical parameters for clusters
were determined in the following way. X-ray luminosities in the 0.1 to
2.4 keV energy band were derived within a cluster radius of 
$r_{500}$ \footnote{$r_{500}$ is the radius where the average
mass density inside reaches a value of 500 times the critical density
of the Universe at the epoch of observation.}. To estimate the cluster
mass and temperature from the observed X-ray luminosity, we used the 
scaling relations described in \citet{Pra2009}. They were
determined from a representative cluster sub-sample of our survey,
called {\sf REXCESS} \citep{Boe2007}
which was studied in detail with deep XMM-Newton observations. Since the radius 
$r_{500}$ was determined from the cluster mass, the calculation of 
X-ray luminosity inside $r_{500}$, the cluster mass, and temperature
were performed iteratively, as described in \citet{Boe2013}. 
The definitive identification of the clusters and the redshift 
measurements are described in \citet{Guz2009}, 
\citet{Cho2012}, and \citet{Boe2013}.

The survey selection function was determined as a function
of the sky position with an angular resolution of one degree
and as a function of redshift. The selection function
takes all the systematics of the RASS exposure distribution, galactic
absorption, and the detection photon count limit into account.
The interstellar hydrogen column density for these calculations is
taken from \citet{Dic1990}.
The selection function as a function of the sky position and redshift was 
published for {\sf REFLEX II} in the online material of \citet{Boe2013}
and for  {\sf NORAS II} in \citet{Boe2017} .

\section{Method}

We studied the density distribution of clusters and of the underlying 
matter distribution as a function of redshift in different regions
of the sky. Because we used a flux-limited cluster sample with 
additional smaller sensitivity variations in regions of the sky with shorter 
exposures, the survey selection function
has to be taken into account to derive the cluster density distribution.
In the following, we describe the method used to correct for 
the selection effects.

We assigned weights to each cluster to correct for
the spatially varying survey limits. The weights were calculated
from an integration of the luminosity function, $\phi(L_X)$, 
as follows:

\begin{equation}
w_i = {\int_{L_{X_0}}^{\infty} \phi(L) dL \over \int_{L_{X_i}}^{\infty} \phi(L) dL} ~~~, 
\end{equation}

where $L_{X_0}$ is the nominal lower limit of the sample 
and $L_{X_i}$ is the lower X-ray luminosity limit at the 
sky location and redshift of the cluster. The adopted X-ray 
luminosity function was determined 
in \citet{Boe2014}. To calculate the
cluster density, we summed up the weights for each cluster
involved, which provides the estimated cluster density
for a volume-limited survey with a lower luminosity limit 
of $L_{X_0}$. This method was used to produce the cluster
density maps on the sky in the next section.
We use a value of $10^{42}$ erg s$^{-1}$
for $L_{X_0}$ throughout the paper.

\begin{figure}[ht]
   \includegraphics[width=\columnwidth]{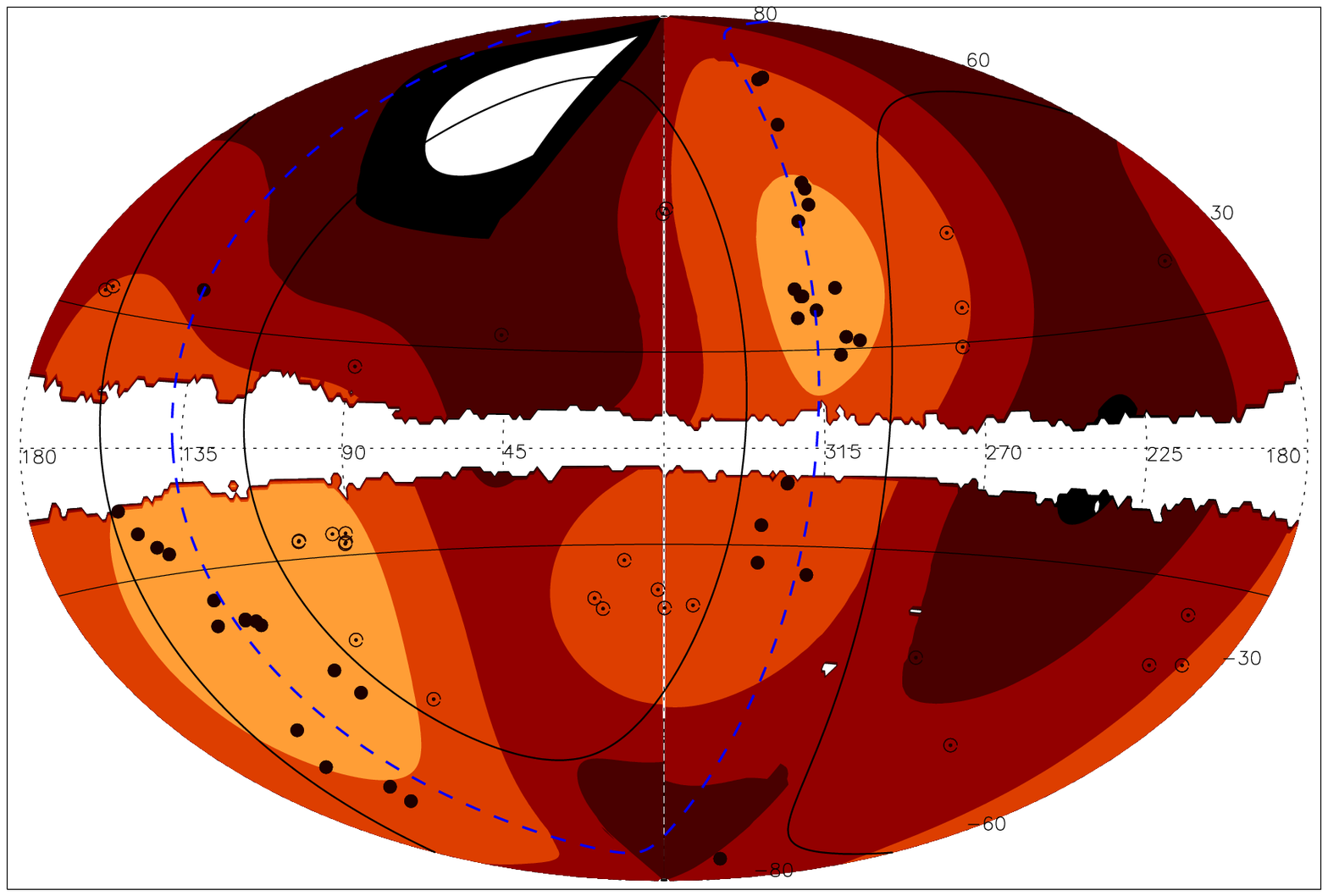}
   \includegraphics[width=\columnwidth]{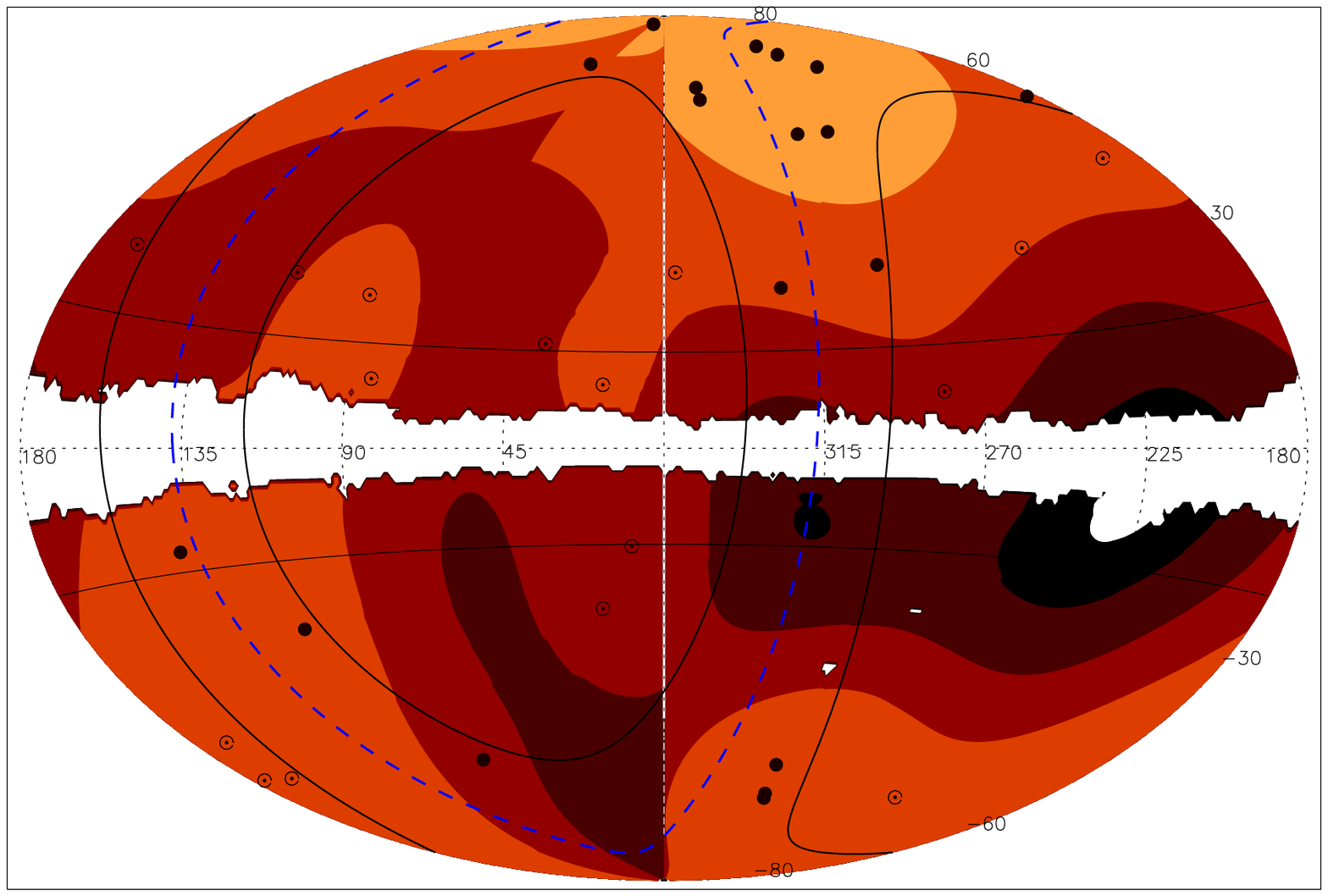}
   \includegraphics[width=\columnwidth]{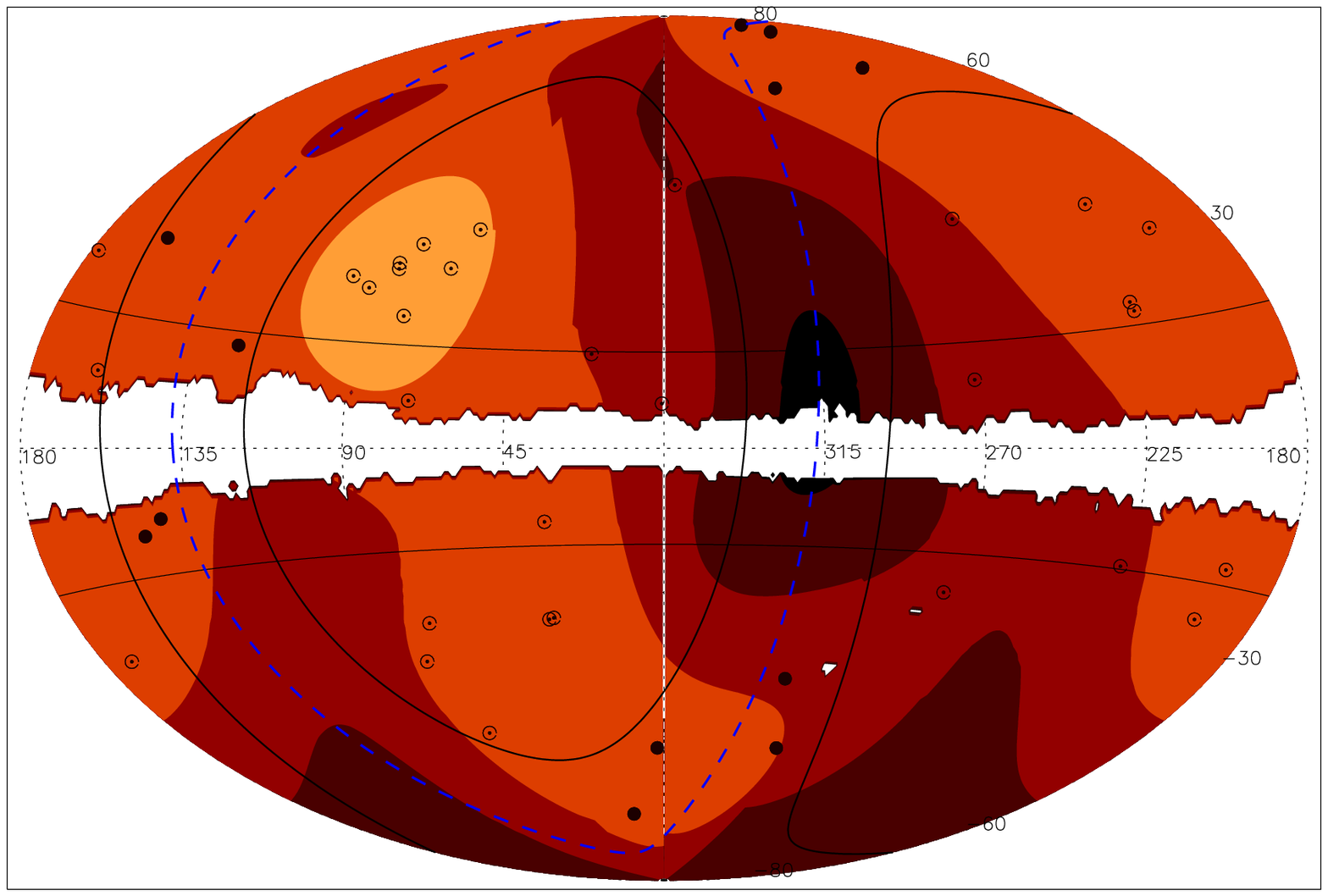}
\caption{Sky distribution of the {\sf CLASSIX} galaxy clusters in
the redshift range $z = 0 - 0.02$ ({\bf top panel}),  $z = 0.02 - 0.025$ ({\bf middle}),
and $z = 0.025 - 0.03$ ({\bf bottom}) in Galactic coordinates. Clusters inside the 
Supergalactic band at latitudes $ \le  \pm 20^o$ are shown as filled circles, while
all other clusters are displayed as open circles. 
The solid lines show Supergalactic latitudes  
$\pm 20^o$ and the dashed line shows the Supergalactic equator. 
The coloured map shows the cluster
surface density as explained in the text. 
}\label{fig1}
\end{figure}

\begin{figure}[ht]
   \includegraphics[width=\columnwidth]{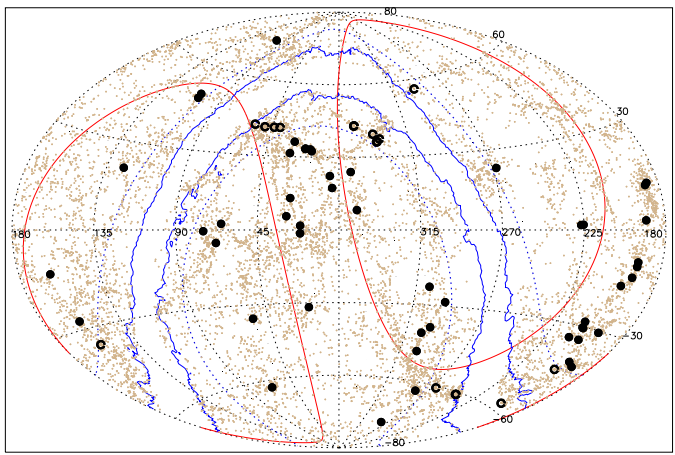}
\caption{Sky distribution of the {\sf CLASSIX} galaxy clusters in
the redshift range $z = 0 - 0.02$ in equatorial coordinates. 
Clusters in the ZoA are shown as open circles, 
while all other clusters are marked with filled circles. The small brown dots show the galaxies
from the 2MASS redshift survey \citep{Huc2012} in the same redshift range. The two black dotted 
lines show Galactic latitudes of $b_{II} = \pm 20^o$, the blue lines show the boundaries
of the zone of high galactic absorption with a hydrogen column density $n_H \ge 2.5 \times
10^{21}$ cm$^{-2}$, and the two solid red lines show the Supergalactic latitudes  
$\pm 20^o$.
}\label{fig2}
\end{figure}

\section{Results}

\subsection{Sky distribution}

A first impression of the distribution of nearby X-ray
luminous galaxy clusters is provided by their location on the sky, 
as shown in Fig.~\ref{fig1}.
The upper panel shows the cluster distribution
out to a redshift of $z = 0.02$ (a distance of about 85.3 Mpc). The 
Supergalactic band with a width of 20 degrees above and below the
Superglatic plane is marked. The region with high galatic absorption 
($n_H \ge 2.5 \times 10^{21}$ cm$^{-2}$) is marked in white in the colour
coding and the few clusters in this region are not considered.
The thin, nearly horizontal lines indicate Galactic latitudes $b_{II} = \pm 20^o$.
In total, 69 {\sf CLASSIX} clusters fall 
into the study region, of which 17 are located in the ZoA.
We used the method described in section 3 to create a 
map of the surface density distribution of clusters smoothed with a Gaussian filter
on a scale of 10 degrees. This map, shown in the figure, was normalised to the 
mean density. Bright areas show overdensities (orange: R > 2, light red: R = 1 - 2, 
where R is the density ratio to the mean density), while dark regions are 
underdense (light brown: R = 0.5 - 1, dark brown and black: R < 0.5).
The bright regions are located in and near to the Supergalactic band.
There is no comparably dense region outside the Supergalactic band.
Most of the clusters are found in two major
concentrations inside the Supergalactic band. These are 
the Perseus-Pisces supercluster and the neighbouring Southern Great Wall in the 
Cetus region (in the lower left quadrant) 
and the Hydra-Centaurus supercluster (around Galactic longitude 315$^o$
and latitude 0 - 40$^o$).

The middle panel of  Fig.~\ref{fig1} shows the {\sf CLASSIX} cluster distribution in the 
next outer shell at $z = 0.02 - 0.025$. The concentration of clusters seen near 
the North Galactic Pole is the part of the Great Wall that falls into this redshift 
shell. This includes the Coma cluster as well as Abell 1367. The regions outside 
the Supergalactic band are mostly occupied by underdense regions. In the next outer 
redshift shell at $z = 0.025 - 0.03$, which is shown in the bottom panel of Fig.~\ref{fig1},
the densest region is no longer close to the Supergalactic plane and major underdense 
regions fall into the Supergalactic band. Thus the segregation of massive structures 
towards the Supergalactic plane does not extend further than a distance of about
$z = 0.025$ ($\sim 106$ Mpc). The concentration of clusters seen in this panel 
consists of groups and clusters not far from the Hercules supercluster, 
which is at a slightly larger distance.

In Fig.~\ref{fig2} we show the {\sf CLASSIX} cluster distribution 
at redshifts $z \le 0.02$ again and compare it to the distribution of galaxies
from the 2MASS redshift survey \citep{Huc2012} in the same redshift region.
We can clearly see that the cluster concentrations coincide with the densest regions
in the galaxy distribution, which is an indication that both are presumably good tracers
of the distribution of dark matter.

\begin{figure}[h]
   \includegraphics[width=\columnwidth]{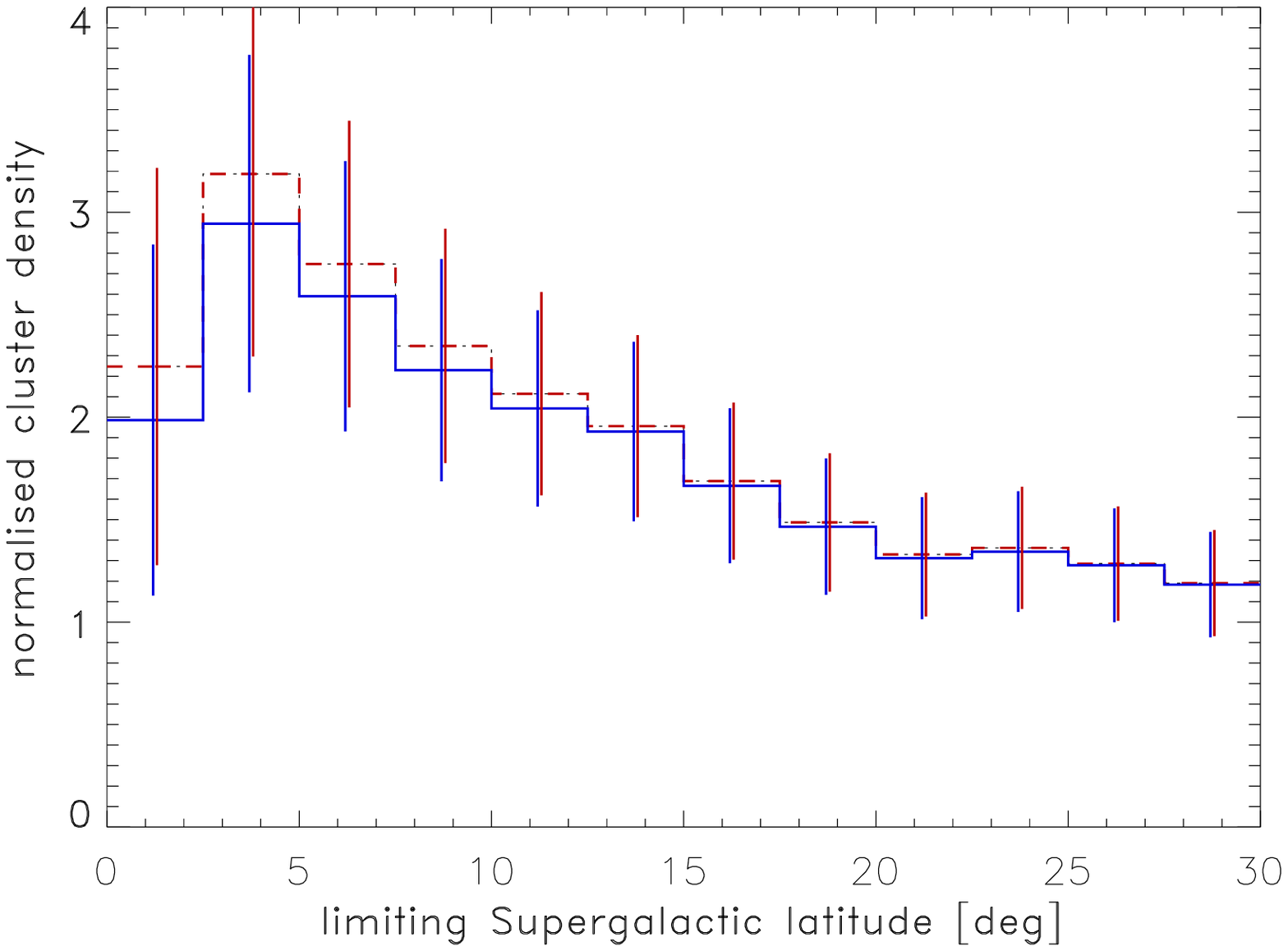}
   \includegraphics[width=\columnwidth]{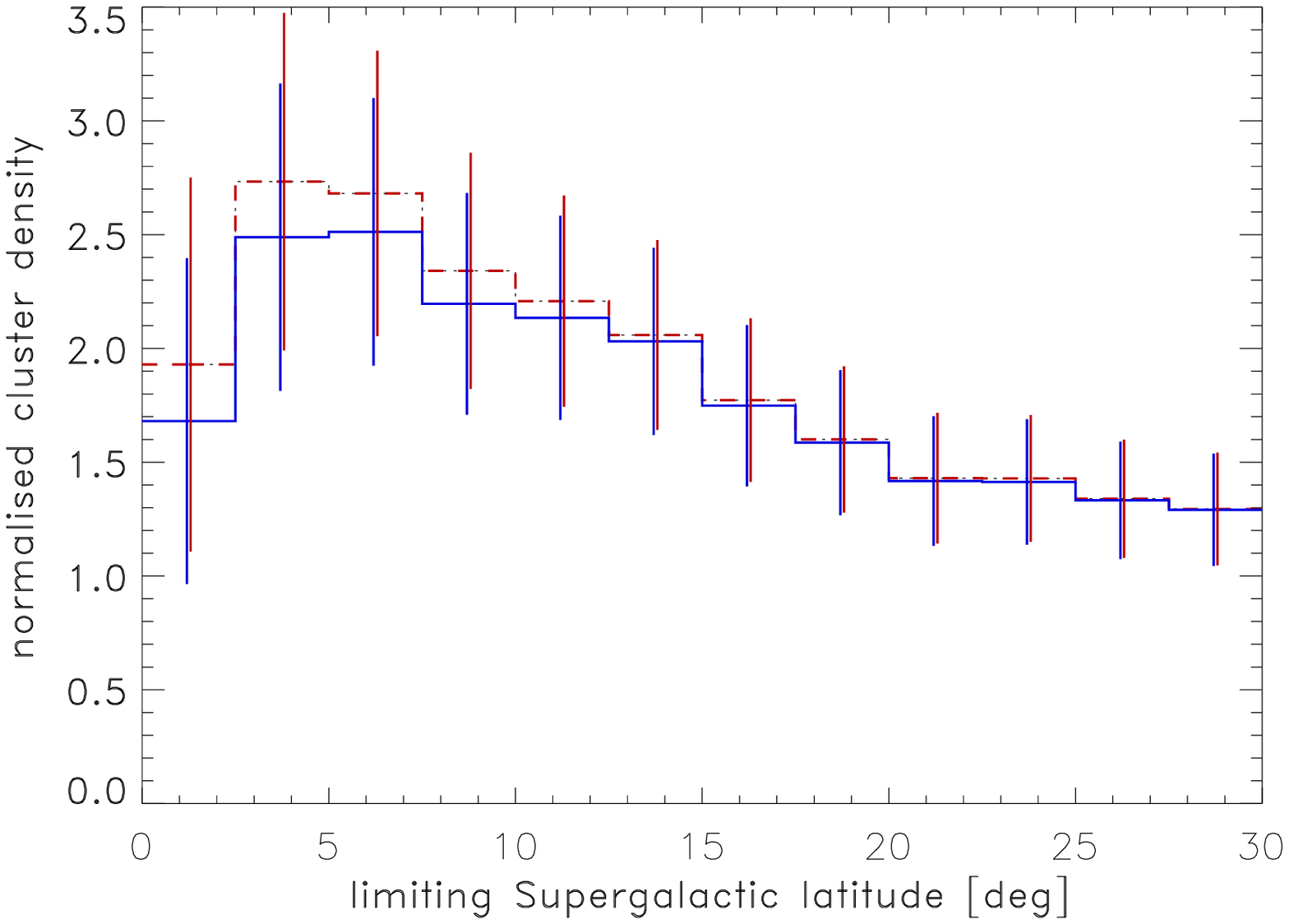}
\caption{Sky surface density ratio of the {\sf CLASSIX} clusters at $z = 0 - 0.02$ 
in a sky area limited by a maximum Supergalactic latitude, $|b_{SG}|$, compared
to the mean surface density in the full volume at $z = 0 - 0.02$.
The solid blue curve shows the density determined with weights correcting for the sensitivity 
variations in the survey as explained in the text, while the dashed red line gives the
unweighted density ratio. The meaning of the error bars is explained in section 4.1. 
The top panel shows the results for the region outside the ZoA, while the bottom panel 
provides the results for the full sky region with hydrogen absorption column density 
$n_H \le 25  \times 10^{20}$ cm$^{-2}$.
}\label{fig3}
\end{figure}

To quantify the concentration of clusters towards the Supergalactic plane, in Fig.~\ref{fig3} we
plotted the surface density of the clusters on the sky as a function
of Supergalactic latitude in the redshift region $z = 0 - 0.02$. Shown is the overdensity 
of clusters in a region for a given maximum Supergalactic latitude, which was determined 
as the ratio of the surface density of clusters in this region to the mean density 
in the full survey area in the given redshift range. 
There is a strong concentration of clusters at latitudes inside 
$\pm 10^o$ with wings of the distribution out to about $\pm 30^o$. The figure
shows the results for two cluster samples. The upper plot only considers the
clusters at  $|b_{II}| \ge 20^o$, where the survey is complete and the selection
function is precise. But this sample covers only 66\% of the sky. In the lower
panel, the survey is extended into the ZoA covering 86.2\% of the sky, but 
the survey completeness and the selection function are not well known in 
the added area, and therefore the result is more qualitative. Nevertheless, 
it shows that there is no prominent structure hidden in the ZoA that would 
significantly change the result shown in the upper panel. 

The error bars shown in the plots are Poisson uncertainties of the cluster number 
counts. We note that these are not to be interpreted as measurement uncertainties 
since the cluster surface density is what was measured directly without errors. If the occurrence
of clusters in this distribution is interpreted as a Poisson point process realisation
of the density values of an underlying density field, then these error bars provide the
uncertainty, with which this underlying density field is represented by the clusters.
In astrophysical terms, it is the accuracy with which the underlying matter density field
can be traced by galaxy clusters. The same interpretation applies to the error bars
shown in Figs. 4 to 6.
 
\begin{figure}[h]
   \includegraphics[width=\columnwidth]{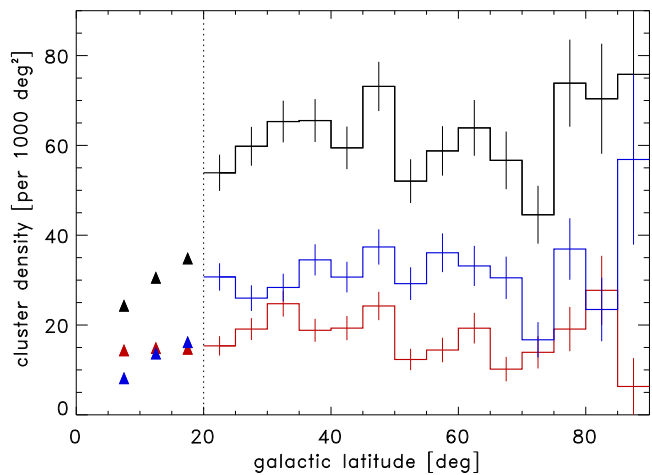}
\caption{Surface density of the {\sf CLASSIX} galaxy clusters on the sky as a function
of galactic latitude, $|b_{II}|$. The black histogram shows all clusters, while the
red one shows clusters in the redshift range $z = 0. - 0.07$ and the blue one clusters
at $z = 0.07 - 0.2$. No clusters are detected closer 
than five degrees to the Galactic plane due to the restriction 
to the region with $n_H \le 2.5 \times 10^{21}$ cm$^{-2}$.
The dotted line shows the boundary of the ZoA. Inside this
boundary, we provide lower limits for the cluster density since  
the redshift survey is incomplete there.
}\label{fig4}
\end{figure}

It is important for this investigation that there is no selection bias of clusters 
as a function of galactic latitude, at least outside the ZoA. 
Thus we show in Fig.~\ref{fig4}, that we find no significant decrease in the surface density with decreasing galactic latitude
for $|b_{II}| > 20^o$. For this study we chose two redshift bins with sufficient 
statistics containing 608 (943) clusters for $z = 0 - 0.07$ ($z = 0.07 - 0.2$),
respectively. The variations in the surface density are larger than the 
error bars, determined from Poisson counting statistics. This should be attributed 
to large-scale structure since we do not see the same variations if we look at
the surface brightness in different redshift shells as shown in Fig.~\ref{fig4}.  
At galactic latitudes $|b_{II}| < 20^o$, we give only lower limits in the figure
since there, in the ZoA, the survey is incomplete. 

\subsection{Radial extent and width of the superstructure}

\begin{figure}[h]
   \includegraphics[width=\columnwidth]{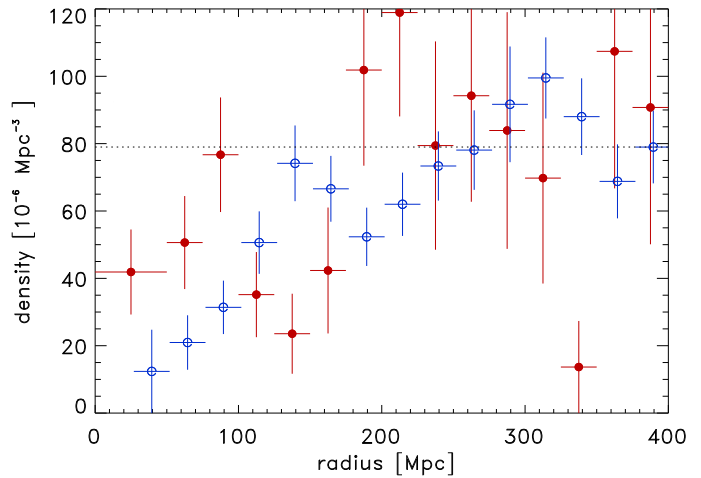}
\caption{Comparison of the density of {\sf CLASSIX} galaxy clusters in
a slice of 50 Mpc width around the Supergalactic plane (red, filled data
points) to the region outside (blue, open data points). Only the volume and
the data outside the ZoA are considered. The densities were determined in
concentric shells with a radial width of 25 Mpc. The cluster densities were 
multiplied with a correction factor to account for survey sensitivity variations. 
The mean density in the survey volume out to 400 Mpc is indicated as a dotted line.
}\label{fig5}
\end{figure}

The next goal is to obtain a quantitative measure of the density contrast
and extent of the overdense structure in three dimensions. 
With this objective, we compared the
cluster density inside a slice of $\pm 25$ Mpc around the Supergalactic plane
to the density outside. To explore the radial extent of the overdensity in the slice,
we divided the space into spherical shells with a radial width of 25 Mpc.
In each shell, we determined the cluster density in the part of the shell which
is contained in the slice as well as the cluster density outside the slice. 
We show the results in Fig.~\ref{fig5}.
For the outer region, there is no contributing volume for the innermost
sphere of 25 Mpc radius and therefore no data point. 
For the inner region, we combined the first two bins
to increase the otherwise poor statistics. The cluster density was 
corrected with cluster weights determined by means of the survey
selection function. The mean density inside the survey volume out
to 400 Mpc is indicated by the dotted line.

We note first that the densities are low at small radii. This result
has been studied in \citet{Boe2020}, where we find a significant
local matter underdensity in the Universe out to a radius of about $170$ Mpc.
Outside this radius, there is a compensating overdensity due
to the presence of several superclusters, the most prominent of
which is the Shapley supercluster. Considering the comparison of the
cluster density in the two local regions, we find that the region close to
the Supergalactic plane has a significantly higher density than the outer region in
all bins inside a radius of 100 Mpc. Outside this radius, the density ratio
reverses for the next two bins. Thus we clearly observe that the flattened
structure of the extended Local Supercluster stretches out to about 100 Mpc
and there is no coherence with this structure at larger radii. 

For the calculation in Fig.~\ref{fig5}, 
we used the data in the volume outside the ZoA.
Including the ZoA in this analysis yields a similar result with a slightly
larger overdensity of the region around the Supergalactic plane.

\begin{figure}[h]
   \includegraphics[width=\columnwidth]{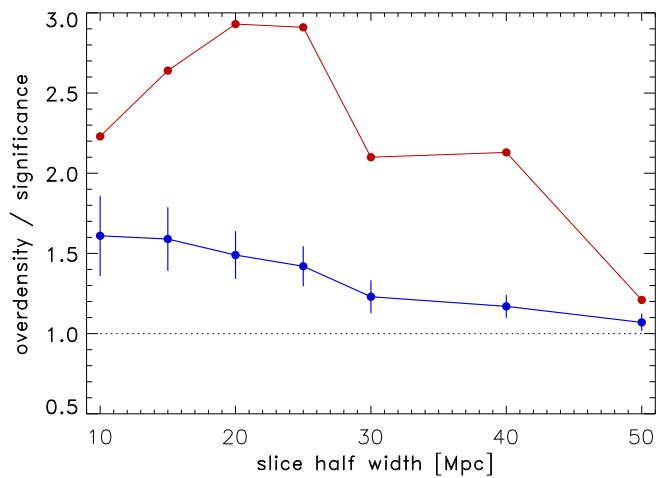}
\caption{Overdensity of galaxy clusters in a slice of half
width as specified on the X-axis with respect to the mean density 
in the survey within a sphere of 100 Mpc radius (lower curve with error bars). 
For the meaning of the error bars, readers can consult the text in sections 4.1 and 4.2.
The upper curve indicates the significance of the 
existence of an overdensity in the underlying matter density distribution.
}\label{fig6}
\end{figure}

To characterise the width of the flattened local 
superstructure, we repeated the study shown in Fig.~\ref{fig5}
for a number of half-width parameters ranging from 10 to 50 Mpc. In all cases, we determined
the overdensity of the flattened superstructure with respect to the mean density
of the full survey volume out to a radial distance of 100 Mpc. The results
are shown in  Fig.~\ref{fig6} where we note an increasing overdensity with a decreasing 
width. The data points directly show the observed cluster density, while the error
bars give again the uncertainty of any implications drawn for the underlying density 
field. The same meaning is also relevant for the significance of the overdensity,
also given in the figure. We note that overall the overdensity and its significance 
is large up to a width of about 50 Mpc, and we adopt this value to best
characterise the extent of the local superstructure perpendicular to
the Supergalactic plane. The significance of the overdensity in the dark matter 
distribution inferred from the cluster sample is about 2.9$\sigma$ (Fig. 6). 

\subsection{Three-dimensional cluster distribution}

\begin{figure}
   \includegraphics[width=\columnwidth]{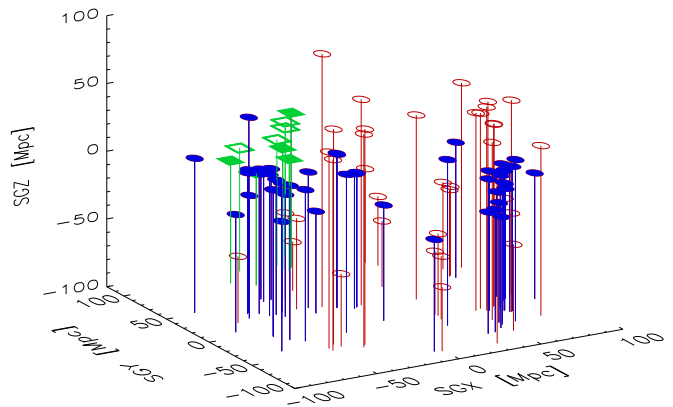}
\caption{Visualisation of the three-dimensional distribution
of the {\sf CLASSIX} galaxy clusters in Supergalactic coordinates
within a radius of 100 Mpc. The clusters within a slice of
50 Mpc width around the Supergalactic plane are marked with filled blue
circles and filled green squares, the clusters outside are shown with red open symbols. 
We note two major cluster 
concentrations associated with the Perseus-Pisces supercluster together with the 
Southern Great Wall (on the right) and Hydra-Centaurus supercluster (on the left).
The cluster members of of the Great Wall in the slice are marked with green 
squares. We also show four further members of the Great Wall inside the slice, but 
just outside the 100 Mpc boundary with green open squares.
}\label{fig7}
\end{figure}

To illustrate the three-dimensional cluster distribution in Supergalactic coordinates,
we used the visualisation presented in Fig.~\ref{fig7}, which shows the {\sf CLASSIX}
clusters used in this study located inside a radius of 100 Mpc. The clusters located in the 
Supergalactic structure in a slice with a width of 50 Mpc are shown with
filled blue circles and filled green squares. 
The remaining clusters are marked with red open symbols. 
We note that most of the clusters are contained in the two concentrations
(as already noted in Figs. 1 and 2) associated with the Perseus-Pisces 
supercluster and the Southern Great Wall
next to it, which appears on the right in Fig.~\ref{fig7}, 
and the Hydra-Centaurus supercluster on the left.

Using a variable linking length with a minimum value of 19 Mpc and a dynamical 
correction using the weighting factors at each
cluster location (a more detailed analysis and description 
will be given in a forthcoming paper),
we constructed memberships to local superclusters. The superclusters found with at least
four members inside a radius of 100 Mpc are Perseus-Pisces with 17 members (of which 13 are located in
the 50 Mpc wide slice and five additional clusters are connected to it outside the 100 Mpc radius), 
the Southern Great Wall with five members (where one cluster is located outside the slice 
and two additional clusters are connected outside 100 Mpc), the Centaurus supercluster with
ten members (all located in the slice), the local concentration 
around Virgo with five members, and parts of the Great Wall. 
Five clusters of the Great Wall are located inside the slice, four clusters are located
outside the slice, and four further clusters are found at a distance between 
100 and 132 Mpc. In Fig.~\ref{fig7} we show the Great Wall members inside the 
slice around the Supergalactic plane with different
symbols (green squares) to better distinguish them from the 
Centaurus Supercluster. In this volume with a radius of 100 Mpc, 52\% of all {\sf CLASSIX}
clusters are contained in the two major superstructures, 
the Perseus-Pisces and Southern Great Wall complex
as well as the Centaurus supercluster. If we add the clusters around Virgo and in
the Great Wall, 70\% would belong to superstructures. More details of these structures and 
their member clusters will be the subject of a forthcoming paper.

The Great Wall appears only partly at the edge of the volume with $r\le 100$ Mpc.
To show the extension of the Great Wall near the Supergalatic Plane, we also 
display the four clusters inside the slice, but  at radii of 100 - 132 Mpc 
with open, green squares.
The local superstructure described by \citet{Tul2014} as the Laniakea supercluster 
comprises the local Supercluster including Virgo, the Hydra-Centaurus supercluster,
and some clusters close to these systems and it makes up most of the concentration seen in
the middle and left part of Fig.~\ref{fig7}.   

\section{Discussion}

We found a strong concentration of the cluster density towards the
Supergalactic plane in a region with a radial extent of about 100 Mpc.
Previously, we found the same region to be underdense with respect to
the mean density of the Universe on
larger scales \citep{Boe2015,Boe2020}. Inspecting Fig.~\ref{fig5},
we note that even though the slice of the local superstructure is locally significantly
overdense, its density is still less than the mean cosmic matter density on large scales.
The inner part of this structure, characterised by Tully and his collaborators  as the 
Laniakea supercluster \citep{Tul2014} has also slightly less than a mean density as 
pointed out by \citet{Cho2015}. In consequence, this large, apparently coherent structure 
is not gravitationally bound and will not collapse in the future to one large massive 
system, but rather disperse into several distinct structures in the frame of $\Lambda$CDM
cosmology \citep{Cho2015}. As part of this structure, our Local Group,
which has a distinct peculiar velocity towards the Virgo cluster, will finally not end
up in a large mass concentration, but it will remain quite isolated when large fractions
of the local superstructure collapse into massive matter concentrations.

For the concentration of clusters towards the Supergalactic plane, we found a density 
contrast of about a factor of 1.5 with respect to the mean density in a slice of half width
of 25 Mpc and a radius of about 100 Mpc (Fig.~\ref{fig6}). 
In projection, the density contrast
in a sky region with a width of about 10-20 degrees around the Supergalactic equator
out to a distance of about 85 Mpc was even larger with a factor of about 2.5 to 3 
(Fig.~\ref{fig3}). The extent of the region in latitude corresponds 
to a width of the slice at $z = 0.02$ of about $\pm 7 - 14$ Mpc.  
The larger number in the latter case is partly due to the measurement within a smaller 
radius and partly due to a more favourable geometry.

In the present study, we excluded the sky region with high interstellar absorption, which
constitutes about 13.8 \% of the sky. In addition the survey in the ZoA is incomplete.
The latter effect should be small at the relevant low redshifts, however, as nearby clusters
are more easily detected and we have given a higher priority to the follow-up of 
low redshift clusters. This is also reflected in Fig. 4, where we can
see no severe deficit of clusters at
low galactic latitudes for the low redshift clusters compared to the high redshift sample. 
Even if we make the extreme assumption that we have missed more than half the clusters 
in the ZoA and that all of these are located outside the local structure, the significance 
for the overdense pancake with a thickness of 50 Mpc would only decrease from 
over 3$\sigma$ to about 2.7$\sigma$. The ongoing eROSITA survey could help to find some of the missing
clusters. But with the above statistics, we do not expect that including all groups and 
clusters possibly hidden in the ZoA would change the conclusion of our study.

Since we have shown with simulations
in our earlier study \citep{Boe2020} that clusters are true tracers of the underlying
matter distribution and that Poisson statistics provides a fair account of the accuracy
with which the density variation can be determined on large scales, we can draw conclusions
on the matter density contrast that is involved in this cosmographical mapping. For
this, we have to take into account that galaxy clusters are biased tracers of the mass distribution
in the sense that the density contrast seen in the clusters is enhanced by a certain bias
factor with respect to the underlying matter distribution. The bias factor can be derived
approximately, as shown in our earlier study \citep{Boe2020}, where we make use
of the theoretical bias relations determined on the basis of large cosmological N-body 
simulations, for example by \citet{Tin2010}. For the lower mass limit of $ 2 - 3.5 \times 10^{13}$ 
M$_{\odot}$ for the {\sf CLASSIX} survey in the redshift range $z = 0 - 0.235$, we estimate 
bias factors of $\sim 1.4 - 1.7$ (for the lower and higher redshift regions,
respectively). The bias of the cluster density is defined as $\Delta_{CL} = 
(n_{CL} - \bar n_{CL}) / \bar n_{CL}) = R_{CL} - 1 = b \cdot \Delta_{DM}$, where
$R_{CL}$ is the cluster density contrast and $\Delta_{CL}$ and $\Delta_{DM}$ 
are the cluster and matter overdensity, respectively.  
Therefore the result implies a density contrast in the matter distribution for the 
local superstructure slice of $R_{DM} =  \Delta_{DM} + 1 = (R_{CL} - 1) / b + 1 $,
which is about 1.3 to 2.3, with $R_{CL}$ taken from Figs.~\ref{fig3} and \ref{fig6},
respectively.
   
\section{Summary and conclusion}

We have shown by means of the distribution of X-ray luminous clusters that the matter 
distribution in the local Universe shows a strong segregation towards the Supergalactic plane.
This segregation is present on small scales of about 6 Mpc in the distribution of local
galaxy groups. It continues in the geometry of the Local Supercluster and extends out to a radius
of about 100 Mpc. At the largest scales, this local superstructure is well traced by 
X-ray luminous galaxy groups and clusters. Perpendicular to the Supergalactic plane,
the structure has a width of about 50 Mpc inside which the cluster density contrast 
is about a factor of 1.5 - 3 compared to the mean. This implies a density contrast in
the matter distribution of about 1.3 to 2.3.

Inside the flattened superstructure, the clusters are mostly concentrated in three
major superclusters: the Hydra-Centaurus complex which has a connection
to the Local Supercluster, the Perseus-Pisces supercluster with a connection to the
Southern Great Wall, and part of the Great Wall.
More than half of the clusters in the local Universe belong to these superclusters.
This is similar to the finding of \citet{Cho2014} studying the fraction of
clusters in superclusters in simulations.
As shown previously, this local region of the Universe is less dense than the global 
mean. Therefore, even if the flat superstructure has a higher density than 
the local mean, it is not dense enough to collapse in the future into one structure
within the standard $\Lambda$CDM cosmological scenario.
It will rather fragment into several pieces, where only the core regions of the superclusters
involved will form the largest future objects.

\begin{acknowledgements} We thank the referee for helpful comments.
We acknowledge support by the DFG through the Munich Excellence
Cluster Universe. G.C. acknowledges support by the DLR under grant no. 50 OR 1905.
\end{acknowledgements}

\end{document}